 \documentclass[floats,aps,          showpacs]{revtex4}

\usepackage{epsfig}

\newcommand {\beq} {\begin{eqnarray}}
\newcommand {\eeqn} [1] {\label{#1} \end{eqnarray}}%

\begin{document}
%
%

\title{Do multineutrons exist?}


\author{
N.\ K.\ Timofeyuk
}                     
%
%

\address{
Physics Department, University of Surrey, Guildford,
Surrey GU2 7XH, England, UK
}

\date{Received: \today}
%

\begin{abstract}
Following   recent work in which events which may correspond 
to a bound tetraneutron  ($^4$n) were  observed, 
it is pointed out  that from the theoretical perspective
the two-body nucleon-nucleon (NN) force
cannot  by itself bind  four neutrons, even if it could bind a dineutron.
Unrealistic modifications of the NN force or introduction
of unreaslistic four-nucleon 
force would be needed in order to bind the tetraneutron. 
The existence of other multineutron systems is discussed.

\end{abstract}

%
\maketitle

 In a recently reported experiment \cite{marques} events were observed 
that exhibit the characteristics of a multineutron cluster liberated in the 
breakup of $^{14}$Be, most probably in the channel $^{10}$Be+$^4$n.
The lifetime of order 100~ns or longer suggested by this measurement, would 
indicate that the tetraneutron is particle stable.  
The existence of a bound tetraneutron, if confirmed,
could challenge our 
understanding of nuclear few-body systems and nucleon-nucleon 
(NN) interactions.
In the present letter, 
I would like to make several observations in this context.
 
{\bf I}

The tetraneutron-like events seen in \cite{marques} deserve attention because
the breakup $^{14}$Be $\rightarrow$ 
$^{10}$Be+$^4$n represents one of the best possible tools to search for
 a tetraneutron.
In earlier  experiments, the tetraneutron was searched for
using  heavy-ion transfer reactions such as 
 $^7$Li($^{11}$B,$^{14}$O)$^4$n \cite{Bel},
$^7$Li($^7$Li,$^{10}$C)$^4$n \cite{Al} and
double exchange  reaction $^4$He($\pi^-,\pi^+$)$^4$n  \cite{U}, \cite{G}.
These reactions require considerable reconfiguring of the target nuclei and
should be strongly suppressed. A
negative outcome from these reactions could be easily anticipated.
On contrary,
the nucleus $^{14}$Be consists of a strongly bound core $^{10}$Be 
and four valence neutrons whose separation 
energy is only  about 5 MeV.
In the attractive field of $^{10}$Be these four neutrons could form
a tetraneutron-like configuration 
which might be shaken off in the $^{14}$Be breakup.

{\bf II}

No proper {\em ab-initio} four-body calculations of the tetraneutron
with
realistic two-body and three-body NN forces are known to the author. However,
several  calculations of the tetraneutron are available.

$(i)$ In Ref. \cite{bev} the tetraneutron,
studied  in the translationally invariant 
  $4\hbar\omega$ oscillator shell model,
was found to be unbound by 18.5 MeV. However, the oscillator
basis is not appropriate for the description of unbound systems.

$(ii)$
No bound tetraneutron was found in Ref. \cite{gorb}  within the
angular potential functions method with semirealistic NN interactions.
No search for the four-body resonance state was carried out.

$(iii)$ 
No bound tetraneutron was found in Ref. \cite{varga} within the
stochastic variational method on a correlated Gaussian basis
for a range of simple effective NN
potentials. No search for the four-body resonance state was
carried out.

$(iv)$ A search for   four-body resonances in the lowest order of the
hyperspherical functions method (HSFM) gave a null result \cite{BBKE},
\cite{SRV}.

$(v)$ The energy behaviour of the
eigenphases, studied in Ref. \cite{GNO} within
the HSFM using  the $K_{max}$ = 6 model space, led the authors to 
the conclusion
that the tetraneutron may exist as a resonance in the four-body
continuum at  an energy between  1 and 3 MeV.
However, a clear indication of the resonance has been seen with only
one of the NN potentials used in the calculations,
namely with the Volkov effective NN force V1 \cite{volkov}.
Volkov effective NN
interactions reproduce the experimental binding energy of another
four-body system, $^4$He, but they bind a dineutron. 
In the particular case of V1, the dineutron is bound by 0.547 MeV.
Therefore, V1 cannot be used in  calculations of  multineutron systems.
Another NN potential, used in Ref. \cite{GNO}, namely that
of Reichstein and Tang  (RT) \cite{RT}, reproduces the n-p triplet
and p-p singlet scattering  lengths and does not bind a dineutron. 
With this potential the energy derivatives of the eigenphases
 monotonically decrease with increasing energy.
This means that
resonances in the four neutron system are absent, at least within the
model space considered.
Since no convergence of the eigenphases with an
increase of the hyperangular momentum has
been achieved in these calculations,
such a conclusion does not look convincing.

To understand whether the RT potential can produce any resonance
or bound state
if the model space of the HSFM is increased,
I have studied the tetraneutron within the extreme uncoupled
adiabatic approximation of the HSFM \cite{fabre} which provides a
lower limit for the binding energy \cite{coelho} within the model space
considered. The model space in these calculations has been increased 
up to $K_{max}$ = 16. In this limit it is sufficient to diagonalise
the matrix of the hyperradial potentials $V_{K'\gamma',K\gamma}(\rho)$,
which include both nuclear and centrifugal force,
and to solve a Schr\"odinger-like equation with the lowest
diagonalised potential $V_{diag}(\rho)$ \cite{fabre}.
The hyperradial potentials have been calculated 
using the technique developed in Ref. \cite{T}.

  The RT potential has only central triplet and singlet even components
$V_{TE}$ and $V_{SE}$ and
the singlet and triplet odd potentials are usually obtained from them
using  an arbitrary parameter $u$ so that
$V_{SO} =  (u-1) V_{SE}$ and $V_{TO} =(u-1)V_{TE}$.
The present calculations have been performed
with $u = 1$ which corresponds to  no interaction in the
odd partial waves. These calculations provide a lower limit for the
binding energy because realistic values of $u$, which lie
in the range of 0.93 to 0.98, induce repulsion between some of neutrons, 
which can only decrease the binding.

The diagonalised hyperradial potentials $V_{diag}(\rho)$,
calculated for different model spaces $K_{max}$,
are plotted in Fig.1a as a function of hyperradius $\rho$.
They have almost converged and are purely repulsive,
monotonically decreasing with the hyperradius,
without any sign of local attractive pockets.
Therefore,   the RT potential can neither
bind the tetraneutron   nor produce any resonances.
To get a bound tetraneutron, the value of $u$ should be increased   to
$u=2.3$. Such a value corresponds to a unphysical attraction
in odd partial waves 
and binds the isotopes $^4$H and $^5$He 
 with respect to the t + n and $^4$He + n thresholds
by about 6.5 MeV and 32 MeV respectively.
In reality, these nuclei are unbound by  3 and 0.9 MeV.

\begin{figure}[t]
\centerline{
        \psfig{figure=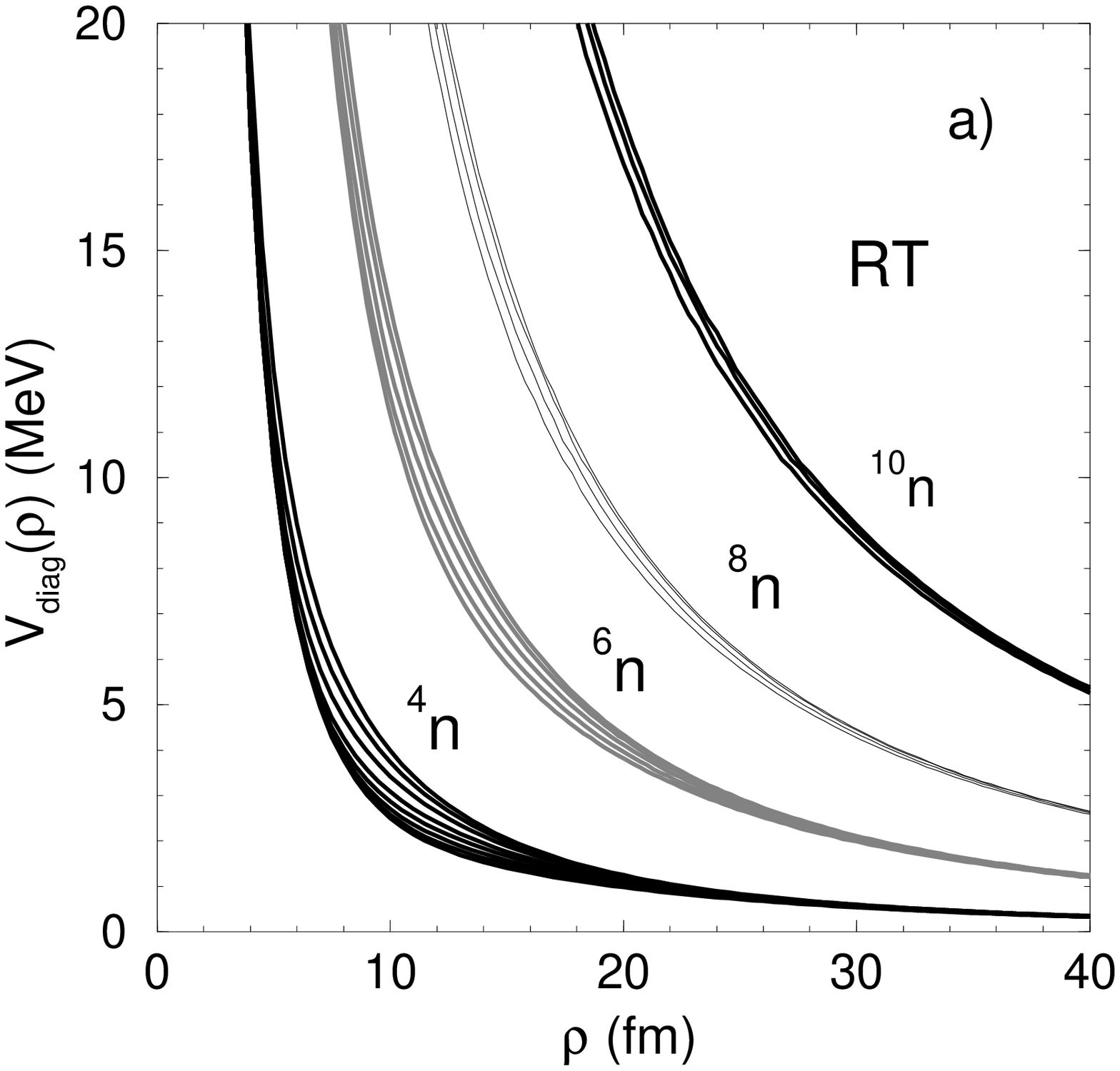,width=0.47\textwidth}
        \psfig{figure=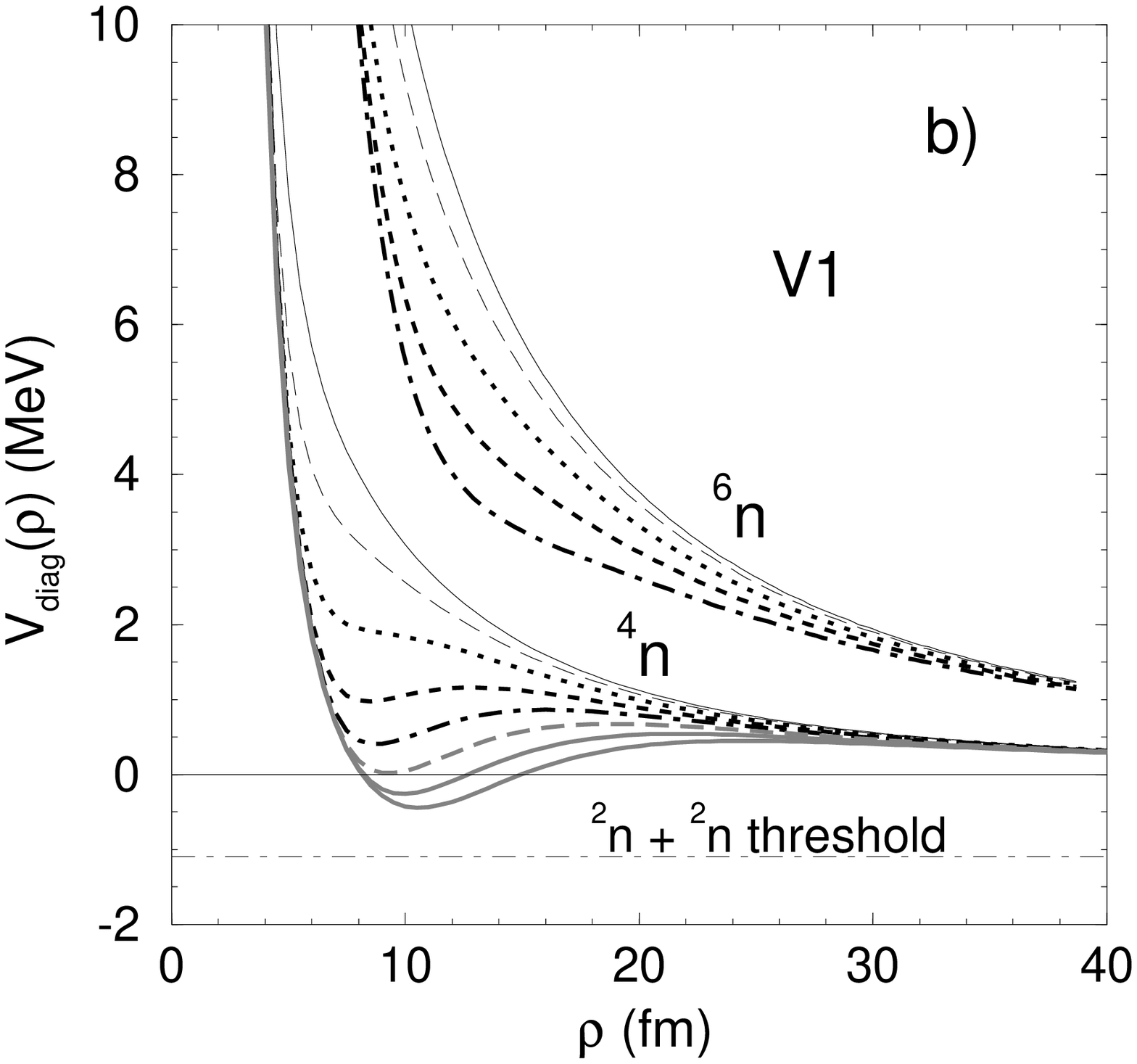,width=0.47\textwidth}}

\caption{$(a)$ The lowest diagonalised hyperradial potentials $V_{diag}(\rho)$
of $^4$n, $^6$n, $^8$n and $^{10}$n
 calculated with $K_{max}$  = 2, 4, ..., 16, $K_{max}$  = 4, 6,..., 12,
 $K_{max}$  =  6,..., 12 and $K_{max}$ = 10, 12, 14
 respectively for the RT  potential; $(b)$ the same as in ($a)$ only
 for $^4$n and $^6$n and the V1 potential.}
\end{figure}

For comparison, the   calculations of the hyperradial potentials
have been performed with Volkov NN potential V1.
The diagonalised hyperradial potentials $V_{diag}(\rho)$,
 shown  in Fig. 1b,  reveal
local    attractive pockets  for $K_{max} > 6$. These pockets
become negative for $K _{max} > 12$, but they are too  shallow
to form a  bound state.
 To get a bound tetraneutron,
the Majorana parameter $m$ should be changed from its standard value of 0.6
to $m = -0.2$. 
Such a change  provides E($^4$n) = $-$1.2 MeV and
does not influence the binding energy of $^4$He, however, it
binds $^4$H and $^5$He with respect to the t + n  
and $^4$He + n thresholds by  7 MeV and 25 MeV respectively.

{\bf III}

The results of the theoretical calculations suggest that  the tetraneutron
must be unbound and most likely should not exist as a resonance,
at least when only two-body central forces are considered.
It is remarkable that even when an effective NN potential
binds the dineutron (the case of V1), it still cannot bind
two  dineutrons, although a resonance in such a system should exist.
In order for the two-body central force  to be able to bind the tetraneutron,
a huge unphysical attraction  must be introduced in  the triplet
odd potential. Such an attraction would strongly overbind $^4$H, $^5$He
and other known $A > 4$ nuclei. 
The   necessity for such a huge attraction lies in the relative
number of nucleon pairs in even and odd states as determined  by the Pauli 
principle.
The NN force binds two nucleons only in the triplet even state. It is
not sufficiently attractive in  
singlet even states,  and   is repulsive in the
triplet and singlet odd states. The probabilities $P$ to find a pair of 
nucleons in all these states are shown in Table 1 for the dominant
components $[22]^{51}S$, $[31]^{33}P$ and $[4]^{11}S$ of
$^4$n, $^4$H and $^4$He  respectively, 
together with binding energies of $^4$H and $^4$He.
One can see that  the binding energy of a nucleus is strongly
correlated with the probabilities to find a pair of nucleons in even states,
especially in the triplet even state where the $n-p$ bound state exists.
In $^4$H, a decrease 
of the probabilities $P_{TE}$ and $P_{SE}$
by 30$\%$ in comparison
with $^4$He  leads to a dramatic decrease of its binding energy by about
20 MeV.
In the tetraneutron, triplet even components (the only ones which produce
binding in the NN system) are absent and the probability to find two neutrons
in the repulsive triplet odd state
increases by 50$\%$ in comparison with $^4$H. Therefore, it is not
unreasonable to assume that  
moving from $^4$H to $^4$n one can loose a comparable
amount of binding energy as going from $^4$He to $^4$H. This would definitely
make the tetraneutron unbound.

\begin{table}
\begin{center}
\caption{Binding energies (in MeV) and the probabilities $P$ to find a pair
of nucleon in the triplet even (TE), singlet even (SE), singlet odd (SO)
 and triplet odd (TO) states for different  $A = 4$ isotopes and for 
 $^{6,8,10}$n.\\}
\begin{tabular}{|c|c|cccc|}
\hline
 & Binding energy &$\, P_{TE}\,$ & $\,P_{SE}\,$ & 
 $\,P_{SO}\,$ & 
 $\,P_{TO}\,$ \\
 \hline
 $\,^4$n$\,$ & &  &1/2  &  & 1/2\\
 $\,^4$H$\,$ & -5.2 &  1/3 & 1/3 &   
  & 1/3 \\
 $\,^4$He$\,$ & -28.3 & 1/2 & 1/2 &   &   \\
 $\,^6$n$\,$ & &  &6/15 &  & 9/15\\
 $\,^8$n$\,$ & &  &5/14 &  & 9/14\\
  $\,^{10}$n$\,$ & &  &1/3 &  & 2/3\\
\hline
\end{tabular}
\end{center}
\label{table}
\end{table} 

 {\bf IV}

The conclusion of particle instability of $^4$n made above
has been drawn on the basis of only two-body NN forces.
The presence of the 3N and, possibly, 4N forces
could, in principle,  modify this conclusion.  
The  contribution of the 3N potential to the
 binding  energies of   neutron drops
$^7$n and $^8$n formed in an external field has been calculated with 
 Green's function Monte Carlo Methods in Ref. \cite{pieper}. It ranges
from 1 to 5 MeV.  
A similar  contribution from the 3N
force could be expected in the case of $^4$n. But such a contribution 
will not  significantly lower the hyperradial potential from Fig. 1a 
 to produce
a potential well in which a bound state of four neutrons could be formed.

As for the 4N force,  the results of the {\em ab-initio} calculations of
the  $3 \leq A \leq 8$ nuclei suggest that
either the 4N 
contributions to the observed  binding energies of these nuclei
are smaller than 1$\%$,  or their effects are included in some parts of the
3N forces  \cite{pieper}.  
The latest models of the
3N force  used to calculated the binding energies
of these nuclei
have not yet been tested in the description of  polarization observables
of the low-energy $Nd$ scattering where the contribution of the
3N force is  important \cite{kiev}, \cite{cadman}.
Besides, the latest numerically accurate calculations employing a 3N
force do not reproduce  the proton
analyzing power for the p-$^3$He scattering \cite{viviani}.
The  simultaneous fit of the $Nd$ and p-$^3$He polarization
observables and  of binding energies of the lightest nuclei
may leave  room for a 4N force.

In order to get an idea of how strong 
the 4N force should be to bind the tetraneutron,
the HSFM calculations  of the present paper have been repeated
with the RT potential using a  4N force   simulated
by the  potential
$V_{4N}(\rho) = W_0 e^{-\alpha \rho}$. The values  $\alpha = 0.7, 1.2$
and 1.5 $fm^{-1}$ 
used in these calculations, were the same as the range of the phenomenological
term of the  3N spin-orbit force   introduced in Ref. \cite{kiev}.
The tetraneutron becomes bound if
 the corresponding values $W_0$ are equal to
$-410$, $-1460$ and $-2530$  MeV respectively. These values are two
orders of magnitude larger than the values $-1$, $-10$, and $-20$ MeV obtained
for  the 3N spin-orbit force \cite{kiev}.
For the Volkov potential V1, these strengths are similar:
$W_0$ = $-320$,  $-1565$ and $-2900$ MeV.
If the same 4N force existed in $^4$He,
the binding energy of $^4$He, calculated with V1,
 would be $-88$, $-82$ and $-134$ MeV respectively instead of $-28.3$ MeV,
 which would be impossible not to notice.

{\bf V}

The reported observation of the tetraneutron revives an old
question  "Do multineutrons exist?" 
Theoretical investigations of   multineutrons have been carried
out only for $^6$n and $^8$n.
The 6$\hbar\omega$ 
oscillator shell model, with the same effective interactions that unbind
the tetraneutron by 18.5 MeV, predicts that 
$^6$n is   unbound  only  by 6.5 MeV \cite{Bev2}. 
However, the with the same interaction, the hydrogen isotope $^6$H
should be unbound with respect to the t + n + n + n decay only by 1.3 MeV
and thus should be less unbound than recently discovered isotope $^5$H
\cite{korsh}. It this was true, the $^6$H isotope would have been already
observed. Other calculations of multineutrons, performed within 
the angular potential functions method \cite{gorb}, 
predict that $^6$n together with another multineutron $^8$n are unbound.
No search for six- or eight-body
resonances have been carried out in these works.

In the present letter, the hyperradial potentials $V_{diag}(\rho)$
have been
calculated for $^6$n with the same NN potentials as for $^4$n, for $K_{max}$ 
ranging from  4 to 12. 
The calculations with the RT potential and $u = 1$, shown at Fig.1a, 
imply that $^6$n should be even more
unbound than $^4$n and that it can not exist as a six-body resonance.
The reason for this lies in further decrease of the number
of the singlet even NN pairs, where attraction occurs,
 and increase of the number of repulsive triplet odd pairs (see
Table 1). In the case of the NN potential V1, which binds the dineutron,
the model space is not large enough to make definite conclusions about  
existence of $^6$n. It is possible that further increase of the model space
will create a local attraction pocket where a six-body resonance could be 
formed but it is unlikely that it will lead to a six neutron bound state.
If  repulsion in the triplet odd components 
was replaced
by attraction in such a way that $^4$n is bound ($u = 2.3$ case for the
RT potential and $m = -0.2$ case for V1) then $^6$n would be bound
by about 22 MeV or 50 MeV respectively. This results does not look surprising
because there is more repulsion in $^6$n than in $^4$n and substitution
of the repulsion by attraction should lead to more binding in $^6$n 
than in $^4$n.

Finally, the hyperradial potentials for $^8$n and $^{10}$n have been 
calculated with the RT 
potential up to $K_{max}$ equal to 12 and 14 respectively.
These potentials, shown in Fig.1a, do not exhibit any sign of either
 a bound or  a resonant state which is consistent with further
decrease of attraction and increase of repulsion in these systems demonstrated
in Table 1.

\vskip 0.3 cm
 Summarizing, due to the small probability for a pair of neutrons to be
 in the singlet even state, the two-body NN force
 cannot  by itself bind  four neutrons, even if it could bind a dineutron.
 Unrealistic modifications of the NN force would be needed to bind the tetraneutron.
 Therefore, it is unlikely that the events in the breakup
 of $^{14}$Be  were caused by a formation of a bound tetraneutron. 
 A different explanation should be sought 
 for this experiment and new experiments are needed to clarify this issue.
 As for experimental searches of other multineutron systems,
 the calculations presented here suggest that
 they might be unsuccessful.\\


The support from the EPSRC grant GR/M/82141
is acknowledged. 
I am  grateful to  Prof. R.C. Johnson and Prof. I.J. Thompson
for useful comments concerning my paper.\\

\end{document}